# INNOVATIVE SEISMIC ISOLATION DEVICES BASED ON LATTICE MATERIALS: A REVIEW

*Fernando Fraternali[1], Ada Amendola[1], Gianmario Benzoni[2]*

[1]Department of Civil Engineering, University of Salerno, Fisciano (SA), Italy

[3]Department of Structural Engineering, University of California San Diego, CA, USA

SUMMARY: *This paper reviews recent literature results on the mechanics of structures formed by layers of pentamode lattices alternating with stiffening plates, which can be effectively employed for the development of seismic isolation devices and vibration attenuation tools, with nearly complete band gaps for shear waves. It is shown that such structures, named pentamode bearings, can respond either in the stretching-dominated regime, or in the bending-dominated regime, depending on the nature of the joints connecting the different members. Their response is characterized by high vertical stiffness and theoretically zero shear stiffness in the stretching dominated regime, or considerably low values of such a quantity in the bending dominated regime. Available results on the experimental response of 3d printed models to combined compression and shear loading highlight that the examined structures are able to exhibit energy dissipation capacity and effective damping that are suitable for seismic isolation devices. Their fabrication does not necessarily require heavy industry, and expensive materials, being possible with ordinary 3-D printers.*

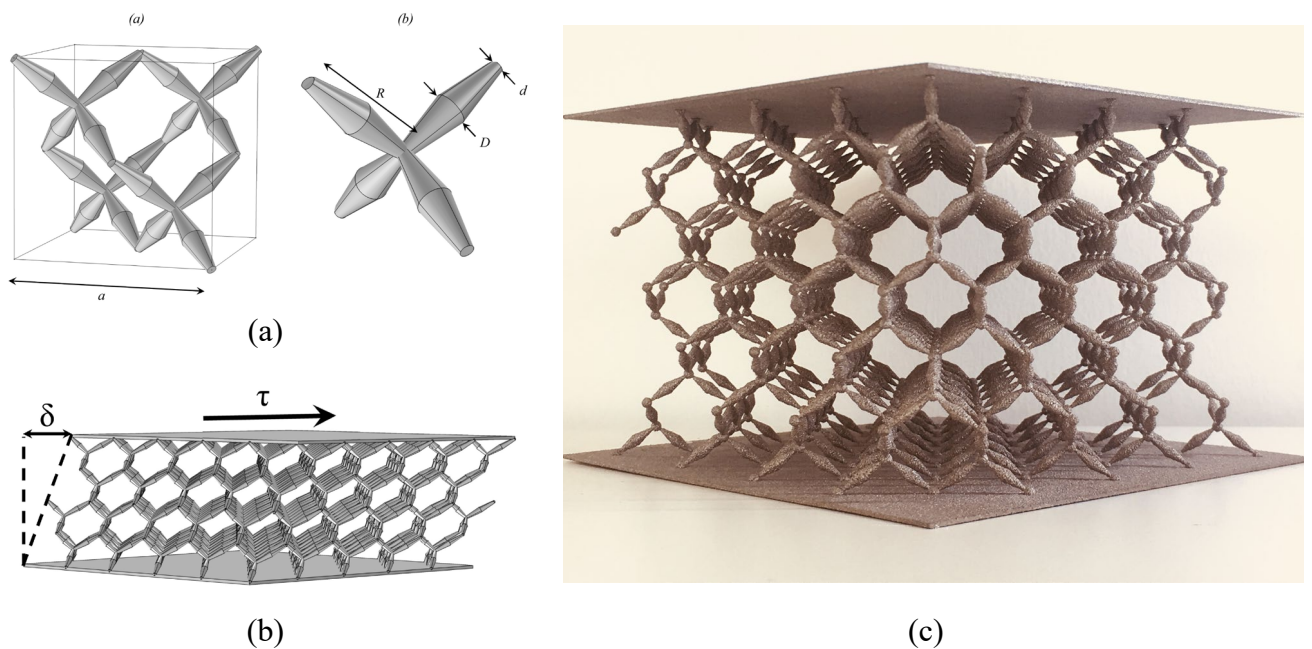

*Illustration of pentamode lattices confined between stiffening plates: (a) geometry of the lattice unit cell; (b) simulation of the response of a single-layer structure under shear loading; (c) 3d printed sample.*

Corresponding author: Fernando Fraternali, Department of Civil Engineering, University of Salerno, Italy.
Email: f.fraternali@unisa.it

# 1. Introduction

Pentamode lattices are 3d lattice structures with elementary unit cell formed by four struts meeting at a point, which are able to exhibit five soft deformation modes and only one stiff mode [Kadic et al. 2013][Milton and Cherkaev, 1995] [Norris, 2014]. The 2d version of such structures is formed by bi-mode metamaterials [Milton, 2013], such as, e.g., honeycomb lattices [Norris, 2014]. Pentamode structures are intensively being studied in different areas of mechanics and physics, due to their capacity of behaving as mechanical metamaterials exhibiting frequency bandgaps for shear waves, and/or elasto-mechanical cloaking [Bückmann *et al.*, 2014] [Chen *et al.*, 2015] [Huang *et al.*, 2016][Martin *et al.*, 2013].

A novel feature of layered structures formed by pentamode lattices and stiffening plates, hereafter referred to as "pentamode bearings" (PMBs), has been recently recognized in the literature, through the discovery that such systems exhibit a three-mode response characterized by stiff response against vertical compression and bending modes, and soft response in shear and torsion [Fraternali and Amendola, 2017] [Amendola et al., 2016b,c,d] [Fraternali *et al.*, 2015]. It is known that unconfined pentamode lattices exhibit (theoretically) zero stiffness against compression loads in the stretch-dominated limit [Milton and Cherkaev, 1995] [Norris, 2014]. The research presented in [Amendola et al., 2016c,2016d] [Fabbrocino et al., 2016] [Fraternali *et al.*, 2015] has revealed that PMBs can instead exhibit high stiffness against compression loading, due to the lateral confinement effect played by the stiffening plates. At the same time, such systems exhibit low (or theoretically zero) stiffness against shear and torsion loads. This kind of response is similar to that exhibited by elastomeric bearings commonly employed as structural supports of buildings and bridge structures [Kelly, 1993], and suggests the use of layered pentamode structures as novel seismic isolation devices [Amendola et al., 2016b, 2016d].

A convenient way of mitigating and potentially eliminate structural and non-structural damages induced by seismic events on buildings and infrastructures is offered by the seismic isolation technique, which essentially consists of increasing the vibration periods of a structural assembly through the insert of a deformable layer obtained, for instance, by the application of isolation devices at the foundation level. This techniques results in the increase of the fundamental structural period to ranges associated to dramatically reduced spectral accelerations [Skinner *et al.*, 1993][Naeim and Kelly, 1990]. Often isolators provides, as part of their performance, a significant energy dissipation capacity otherwise obtained through the use of ad hoc additional devices which are aimed at reducing the amplitude of the lateral displacements of the isolated structure [Naeim and Kelly, 1990].

The capacity to support gravity loads (and so the bearing characteristics) is stressed in the European Standard EN 15129 [EN 15129,2009] where a seismic isolator is defined as: "d*evice possessing the characteristics needed for seismic isolation, namely, the ability to support a gravity load of superstructure, and the ability to accommodate lateral displacements*". Nowadays, two are the most commonly used families of seismic isolators. The first one comprise the elastomeric bearings, a category of structural devices with different energy dissipation capacity ranging from low to high damping and very high dissipation capacity obtained with the lead-rubber bearings [Kelly, 1993][Skinner *et al.*, 1993][Naeim and Kelly, 1990]. All the elastomeric bearings are formed by alternating steel plates (shims) and layers of rubber (natural or synthetic) in order to provide bearing capacity and stability while allowing shear deformations. Top and bottom steel plates are also provided for connection to the existing structure. Lead-rubber bearings add to the previous configuration one or more lead-plugs, characterized by high elasto-plastic dissipation capacity, and inserted perpendicularly to the metal shims. Detailed experimental data about the performance of elastomeric bearings can be found in [Benzoni and Casarotti, 2009].

The second most common category of seismic isolators is that of friction-based devices, in which articulated sliders are allowed to move on spherical stainless steel surfaces. The interface of the moving components is obtained with proprietary low friction materials and stainless steel. The mechanism of energy dissipation is obtained through the frictional features of the interfaces activated during the motion [Lomiento *et al.*, 2013a and 2013b]. The large popularity achieved in the last decade by these devices is due to the very low profile (suitable for retrofit interventions), the small footprint for large displacement capacity and the controlled frequency of the motion function of the surface curvature (pendulum effect). For structural configurations in need of supplemental energy dissipation feasible devices are, e.g., viscous dampers, yielding-based dissipation elements, friction dampers as well as different smart devices employing, e.g., the super-elastic response and the re-centering capacity of shape memory alloys [Pant et al., 2013] [Skinner *et al.*, 1993][Naeim and Kelly, 1990][Graesser and Cozzarelli, 1991].

It must be noted that the above seismic isolator families lack the capacity to sustain a moderate/severe tensile state due to vertical loads. The elastomeric devices in fact, while capable to resist a limited amount of tensile force, rapidly experience delamination and failure. The friction based devices instead can be treated as articulated assembly with zero tensile capacity. When this scenario is realistic, like due to overturning behavior of the structure, restraining additional devices could be required. For critical structures, like power plants, solutions of devices with isolation capabilities in both horizontal and vertical directions are still under development.

The literature results reviewed by the present study highlight that PMBs can be designed to behave as tension-capable and performance-based isolation devices, whose properties mainly depend on the geometry of the members forming the unit cell of the lattice, the nature of the connections joining the strut one another, and the struts and the stiffening plates, as well as the mechanical properties and sizes of the component materials [Fraternali and Amendola, 2017] [Amendola et al., 2016b,c,d] [Fraternali et al., 2015]. [Amendola et al., 2018, 2018a]. The response of such devices can be easily adjusted to that of the structure to be protected, and their fabrication does not necessarily require heavy industry, and expensive materials, being possible with ordinary 3-D printers.

## 2. Layered pentamode structures

The present study is focused on layered structures formed by alternating layers of pentamode lattices, which may exhibit different geometries and numbers of unit cells, and stiffening plates. A first set of structures that we examine hereafter are the so-called *fcc lattices*, which obtained by tessellating the 3d space with the usual, face-centered-cubic (fcc) pentamode unit cell (see Figure 1 (a)). A second set of structures is instead obtained by tessellating in the horizontal plane the subset of the fcc unit cell shown in Figure 1 (b), which we name sfcc unit cell (*sfcc lattices*).

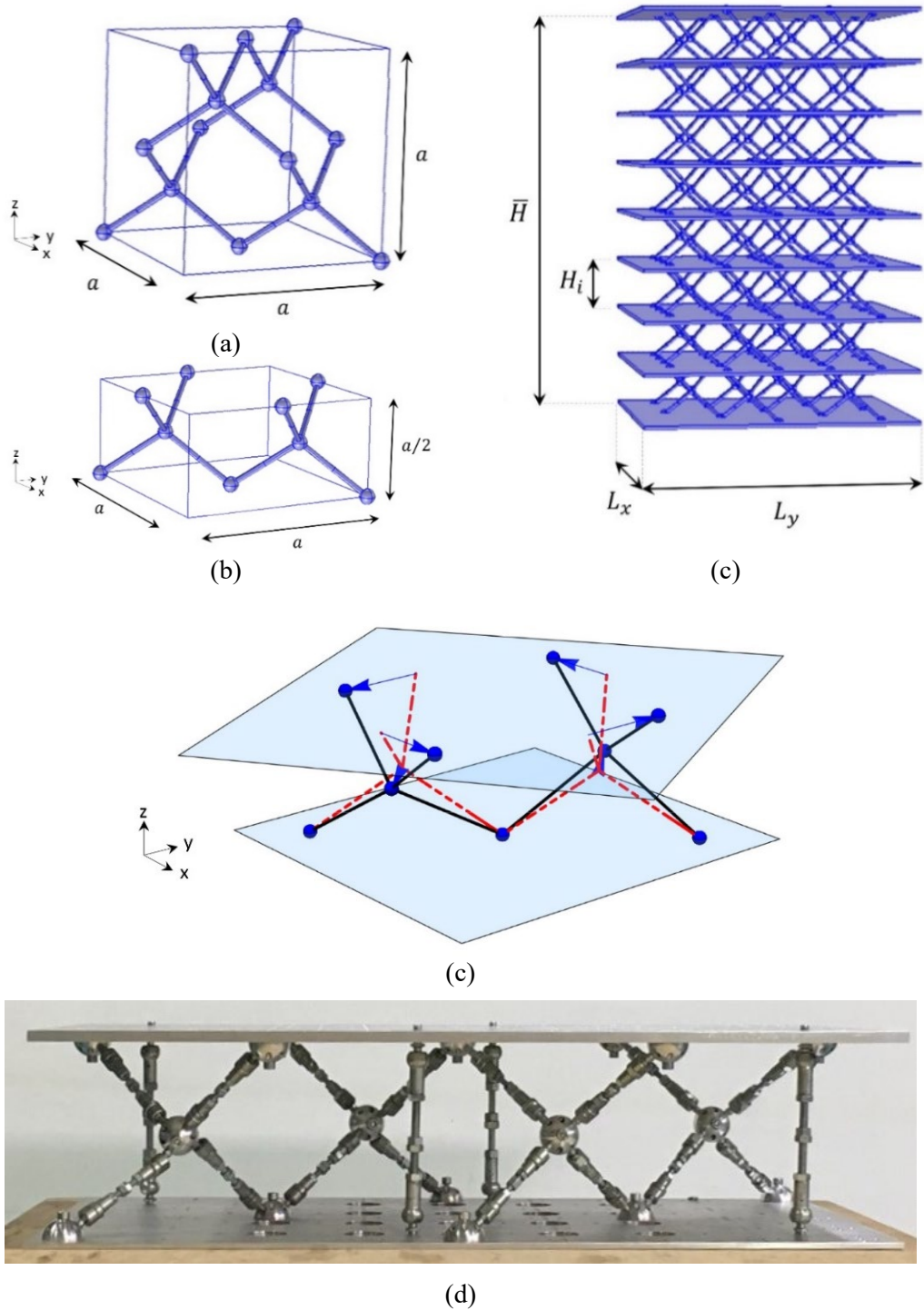

Figure 1 - *(a) fcc unit cell formed by rods with uniform cross-section; (b) sfcc unit cell; (c) layered structure equipped with hinged connections; (c) infinitesimal mechanism of the unit cell under torsion; (d) physical model including supplementary vertical rods (cf. Sect.3).*

For what concerns the connections between the struts and the joints between the struts and the stiffening plates, we examine two different cases: a first case in which all the joints are hinged/pinned and are made, e.g, of the hollow ball joints commonly used in structural space grids [Chilton, 2000] (Figure 1); and a second case in which all the connections forming the structure are rigid/moment resisting (Figure 2). In the first case, we assume that the struts have uniform cross-section (Figure 1), while in the second case we account for rods that have a bi-conical profile, whose radius progressively shrinks when passing from the mid-span to the joints, so that we can be able to analyze different values of the bending rigidity of the nodes [Gurtner and Durand, 2014] [Norris, 2014] (Figure 2).

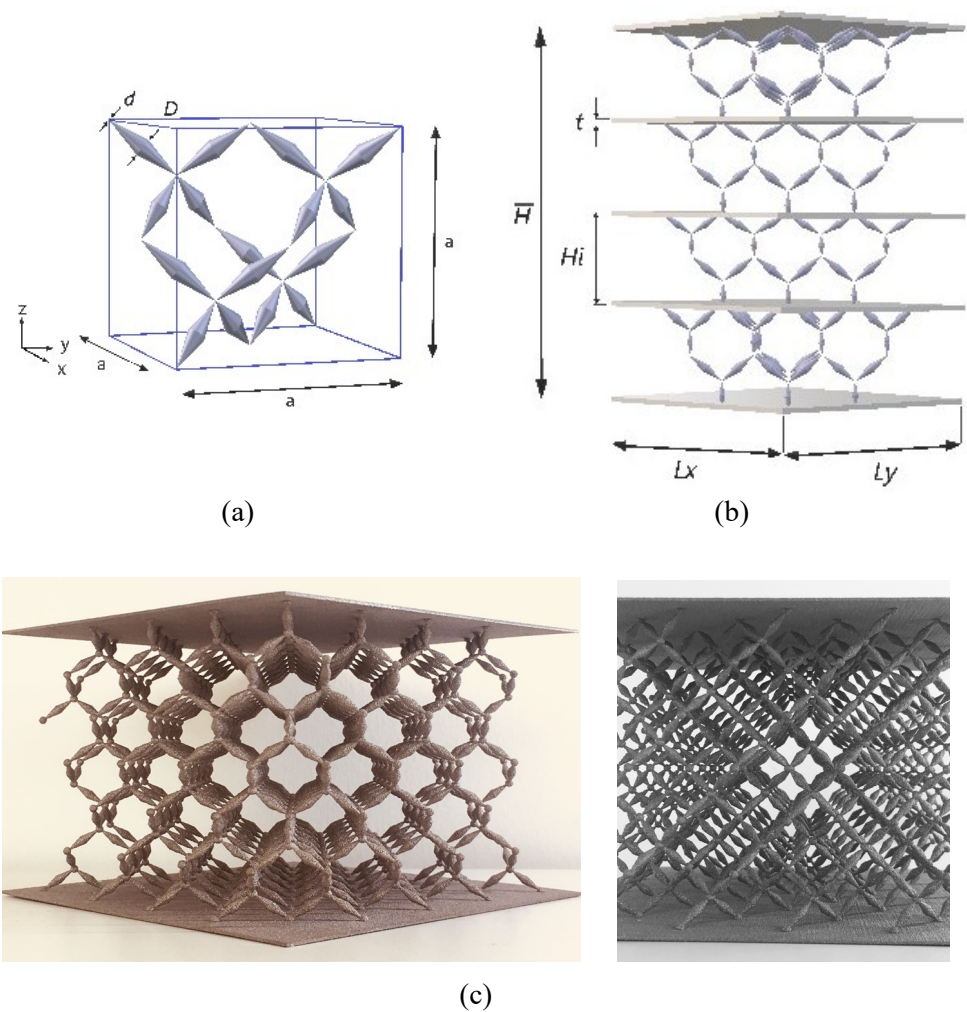

Figure 2 - *(a) fcc unit cell formed by bi-conical rods with variable cross-section; (b) fcc layered structure featuring rigid connections; (c) 3d printed samples (cf. Sect. 4).*

Referring to Figure 1 and Figure 2, the following makes use of the symbol $a$ to denote the lattice constant (characteristic size of the unit cell); the symbols $L_x$, $L_y$ to denote the edge lengths of the stiffening plates, and the symbols $n_x, n_y$ and $n_z$ to denote the number of unit cells placed along the $x, y$ and $z$ axes of a Cartesian frame aligned with the edges of the unit cell ($z$ axis placed along the vertical direction). The thickness of the generic layer is denoted by $H_i$; the total thickness of the pentamode layers is denoted by $H = n_z H_i$; and the total height of the laminated structure (including the stiffening plates) is denoted by $\bar{H}$.

## 3. Pure stretching regime

We begin by examining the sfcc layered structures illustrated in Figure 1, letting $s$ denote the cross-section area of the rods, and letting $E_0$ denote the Young modulus of the material forming the rods. The results presented in the current section assume that the stiffening layers respond as rigid plates, and the pentamode layers respond in the pure stretching regime induced by the presence of hinged connections in correspondence of all the joints of the structure.

The analytic study presented in Ref. [Fraternali and Amendola, 2017] has obtained analytic formulae for the vertical stiffness $K_v$ and bending stiffness $K_\varphi$ coefficient of a laminated sfcc structure formed by an arbitrary number of layers. Such formulae read as follows

$$K_v = \frac{1}{\sum_{i=1}^{n_z} \frac{1}{K_{v_i}}}, \quad K_\varphi = \frac{1}{\sum_{i=1}^{n_z} \frac{1}{K_{\varphi_i}}}, \tag{1}$$

where $K_{v_i}$ and $K_{\varphi_i}$ are the vertical stiffness and the bending stiffness (about either the $x-$ or the $y-$ axis) of the generic layer. The latter are given by

$$\frac{K_{v_i}}{E_0 a} = \frac{2}{9} n_x n_y \phi, \quad \frac{K_{\varphi_i}}{E_0 a^3} = \frac{1}{216} n_x (4 n_x^2 - 1) n_y \phi, \quad \phi = \frac{8 s \ell}{a \times a \times \frac{a}{2}} = \frac{4\sqrt{3} s}{a^2}, \tag{2}$$

$\phi$ denoting the solid volume fraction of the unit cell. The effective compression modulus of the layered structure is given by

$$E_c = \frac{K_v \frac{a}{2}}{n_x n_y a^2} = \frac{4 E_0 s}{3\sqrt{3} a^2} = \frac{\phi}{9} E_0. \tag{3}$$

The shear and twisting stiffness coefficients (as well as the effective shear modulus $G_c$) are instead equal to zero in the pure stretching regime under examination.

It is interesting to note that, while the compression modulus of an unconfined pentamode lattice is zero in the stretch-dominated limit [Norris, 2014], Eqn. (3) reveals that the effective compression modulus of a confined sfcc structure is equal to 2/3 of the Young modulus of the stiffest isotropic elastic networks analyzed in Ref. [Gurtner and Durand, 2014]. The same equations also shows that $E_c$ increases with the rods' cross-section area $s$, and decreases with the lattice constant $a$, being a linear function of the solid volume fraction $\phi$.

The compression modulus $E_c$ of an elastomeric layer confined between stiffening plates, which is employed in rubber bearings, is ruled by a shape factor defined as the ratio between the load area and the force-free (lateral) area [Kelly, 1993][Benzoni and Casarotti, 2009][Skinner et al., 1993][Hutchinson and Fleck, 2006]. Such a quantity is hence directly proportional to a characteristic dimension of the load area, and inversely proportional to the rubber pad thickness (see, e.g., [Skinner et al., 1993]). Eqn. (3) points out that the compression modulus of a sfcc

PMB is inversely proportional to the lattice thickness ($a/2$, cf. Figure 1b). One notes that the role played by the characteristic transverse dimension of the rubber pads in rubber bearings is replaced by the cross-section area of the rods in a PMB.

It is now instructive to examine the special case of a square system such that it results $n_x = n_y = n_a$, and one can write

$$a = \frac{L}{n_a} = \frac{2H}{n_z}, \tag{4}$$

with $L = L_x = L_y$. The theory presented in Ref. [Fraternali and Amendola, 2017] leads to the following results in the case under examination

$$K_v = \frac{2}{3\sqrt{3}} \frac{E_0 \, s \, L}{H^2} n_a n_z \tag{5}$$

$$E_c = \frac{2}{3\sqrt{3}} \frac{E_0 s}{LH} n_a n_z \tag{6}$$

Eqns. (5) and (6) reveal that $K_v$ and $E_c$ exhibit linear scaling laws with the number of unit cells in the *x-y* plane and the number of layers.

The study presented in [Fabbrocino and Amendola, 2017] has been oriented to apply the analytic results presented in Ref. [Fraternali and Amendola, 2017] to the design of sfcc PMBs featuring the same vertical stiffness $K_v$ of a commercial rubber bearing having diameter 0.85 m and height 0.35 m, which is composed of 29 rubber layers of 7 mm each, and 28 steel plates (or steel shims) of 3.04 mm each, two terminal rubber layers of 31.8 mm each and covers (isolator Type E [RB-800] produced by Dynamic Isolation Systems, Inc, McCarran, NV, USA).

The use of Eqns. (2), (4) and (5) leads us to obtain for a square PMB the following expression of the vertical stiffness coefficient

$$\frac{2}{9} E_0 L \frac{n_a}{n_z} \phi = K_v \tag{7}$$

On the other hand, by letting $f_y$ denote the yielding stress of the material composing the bars and letting $\alpha$ a safety factor, the limit design approach presented in [Fabbrocino and Amendola, 2017] leads us to write

$$\frac{2}{3} E_0 n_a \frac{w}{n_z L} = f_y \alpha \tag{8}$$

where $w$ is the vertical displacement applied to the PMB. Using Eqn. (8) in association with Eqn. (7), we obtain the following design formulae for $n_a$ and $\phi$

$$n_a = \frac{3\alpha f_y n_z L}{2E_0 \dot{w}}, \quad \phi = \frac{3K_v w}{\alpha f_y L^2} \tag{9}$$

Let us now employ steel bars grade S335JH, with Young modulus $E_o = 210$ GPa and yield strength $f_y = 355$ MPa, and let us assume $L = 600$ mm, $w = 1.00$ mm, and $\alpha = 0.5$. Let us also assume $K_v = 733$ kN/mm, as it results in the commercial rubber bearing under consideration [Fabbrocino and Amendola, 2017]. Table 1 and Figure 3 illustrate different PMBs designed according to Eqns. (9), which exhibit different numbers of layers $n_z$. It is seen that the bilayer pentamode bearing of design (c) exhibits horizontal and vertical dimensions rather similar to those of the commercial bearing under examination.

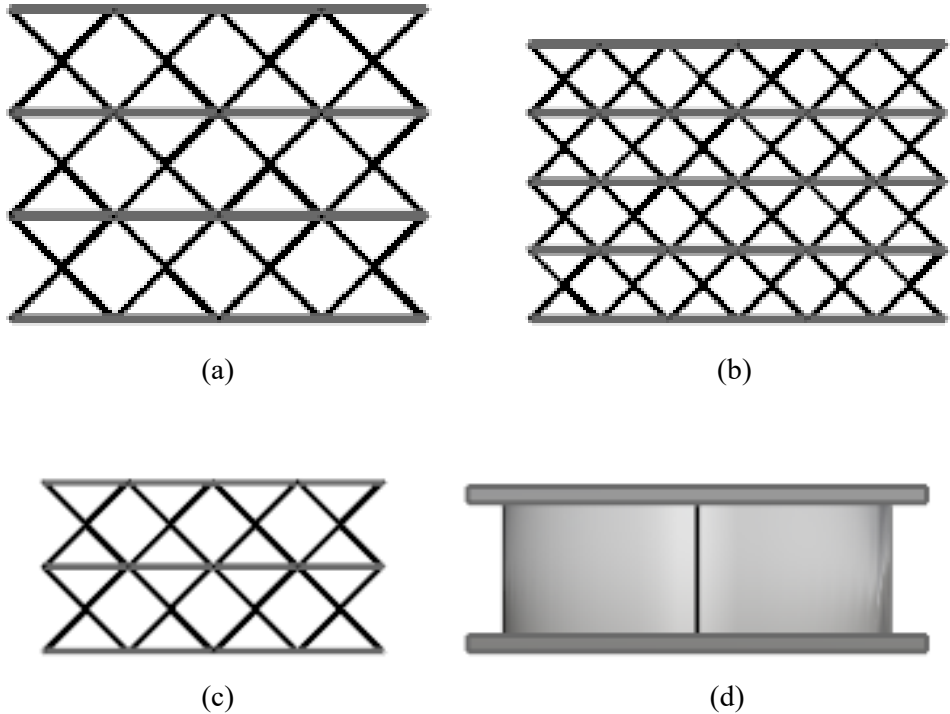

Figure 3 - *Pentamode bearings (a, b, c) featuring equal vertical stiffness and load-carrying capacity of a commercial rubber-bearing (d).*

Table 1 - *Geometrical properties of PMBs replicating a commercial rubber bearing*

|  | $n_z$ | $n_a$ | a [mm] | H [mm] |
|---|---|---|---|---|
| (a) | 3 | 2 | 300 | 450 |
| (b) | 4 | 3 | 200 | 400 |
| (c) | 2 | 2 | 300 | 300 |

A study presented in Ref. [Amendola *et al*., 2016b] extended the above study in order design a PMB entirely made of steel (for both for the stiffening plates and the pentamode lattices), which shows vertical and lateral stiffness coefficients approximatively equal to that of the commercial seismic isolator under consideration. It is worth noting that the nonzero tensile strength of the rods forming the pentamode lattices makes PMBs tension-capable.

One relevant limitation of hinged PMBs is that they behave as finite mechanisms in the large displacement regime and that such mechanisms lead the top-base to progressively collapse on the bottom base [Amendola *et al*., 2016b]. This kind of response is not desired in base isolation systems, since it may lead the isolated structure to be affected by loss of stability, especially in the case of tall buildings. A way to circumvent such an issue is to equip PMBs with rigid or semi-rigid connections [Amendola *et al*., 2016b]. The following sections of this paper examine the bending-dominated response of layered pentamode structures equipped with rigid connections and tapered rods.

## 4. Bending-dominated regime

The bending dominated regime of PMBs equipped with rigid connections has been studied in a series of studies, which have examined the experimental response of additively manufactured reduced-scale samples made of a titanium alloy [Amendola *et al*.,2016d], and the simulated response of numerical models of PMBs for varying aspect rations and lamination schemes [Amendola *et al*., 2016c] [Amendola *et al*., 2018].

The PMBs examined hereafter have the geometry illustrated in Figure 2 and are composed of rods featuring a tapered, bi-conical shape with large diameter *D* at the mid-span and small diameter *d* at the extremities [Amendola *et al*., 2016c,d][Amendola *et al*., 2018].

### 4.1 Physical models of PMBs

The experimental study presented in Ref. [Amendola *et al*., 2016d] has examined 3d printed models of PMBs manufactured through the Arcam S12 Electron Beam Melting (EBM) facility available at the Department of Materials Science and Engineering of the University of Sheffield, using the titanium alloy Ti-6Al-4V (hereafter labeled Ti6Al4V, main properties in Table 2 [Metals Handbook, 1990]). The employed additive manufacturing technology is diffusely described in Refs. [Amendola *et al*., 2016d] [Hernandez et al., 2015] [Van Grunsven *et al*., 2014]

Table 2 - *Main physical and mechanical properties of Ti6Al4V [Metals Handbook,1990]*

| | |
|---|---|
| mass density [g/cm$^3$] | 4.42 |
| yield strength [MPa] | 910.00 |
| Young's modulus [GPa] | 120.00 |
| Poisson's ratio | 0.342 |

The physical models examined in Ref [Amendola *et al.*, 2016d] assume constant lattice constant $a = 30\ mm$ and diameter at the mid-span of the rods $D = 2.72\ mm$ ($D/a \approx 9\%$). A collection of samples have been manufactured by letting the diameter of the rods at the junctions to assume the following different values of $d$: $d_1 = 0.49\ mm$ ($d_1/a = 1.6\%$); $d_2 = 1.04\ mm$ ($d_2/a = 3.5\%$); and $d_3 = 1.43\ mm$ ($d_3/a = 4.8\%$). Table 3 shows the geometric properties of the CAD models of the samples analyzed in Ref. [Amendola *et al.*, 2016d], which do not perfectly match those of the 3d printed models (provided in the same table): This is due to the fact that the EMB process leads to manufacture objects characterized by appreciable surface roughness, which results in larger sizes of the 3d printed objects as compared to the CAD models [Hernandez et al., 2015] [Van Grunsven *et al.*, 2014].The samples corresponding to the smallest diameter of the rods at the junctions ($d_1/a = 1.6\%$) are characterized by less pronounced bending rigidity of the junctions, as compared to the samples characterized by larger values of *d* [Gurtner and Durand, 2014] [Norris, 2014]. The *d/a* parameter defines a microstructure aspect ratio. A macrostructure aspect ratio is given by the ratio between the total height of the sample *H* and the lattice constant *a*.

Table 3 - *Geometrical parameters of CAD and EBM built models of PMBs*

|  | a [mm] | D [mm] | $d_1$ [mm] | $d_2$ [mm] | $d_3$ [mm] |
|---|---|---|---|---|---|
| Built size | 30 | 2.72 | 0.49 | 1.04 | 1.43 |
| CAD size | 30 | 2.71 | 0.45 | 0.90 | 1.35 |

The experimental study presented in Ref. [Amendola *et al.*, 2016d] considers "slender" specimens with *H/a*=4, and "thick" specimens with *H/a*=2. In all such samples, the fcc unit cell id replicated 2x2 times in the horizontal plane, and the stiffening plates feature 80 mm edge and $t_p = 1$ mm thickness. The slender specimens are named SPM1,2,3 when it respectively results $d = d_1$, $d = d_2$, and $d = d_3$ (Figure 4); similarly the thick specimens are named TPM1,2,3 when it results $d = d_1$, $d = d_2$, and $d = d_3$, respectively (Figure 5).

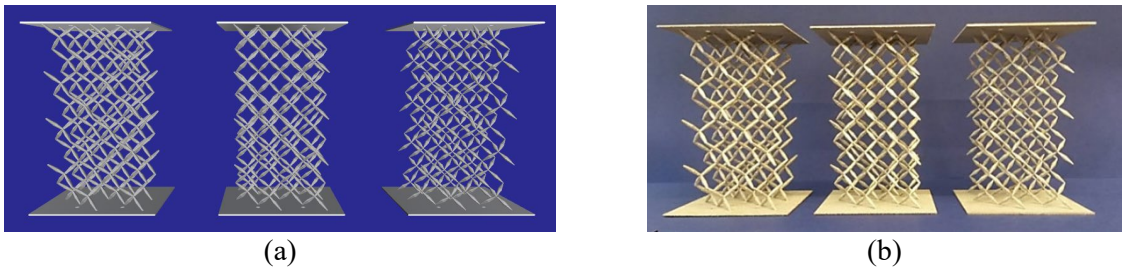

(a)          (b)
Figure 4 - *CAD (a) and 3d printed models (b) of slender specimens.*

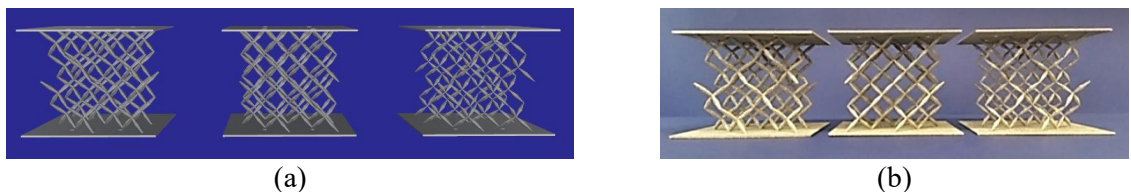

(a)          (b)
Figure 5 - *CAD (a) and 3d printed models (b) of thick specimens.*

## 4.2 Experimental tests

The physical models illustrated in the previous section were subject to lateral force-displacement histories under constant vertical load (hereafter referred to as "shear tests"), first before macro-yielding of the material occurred, and next in the post-yielding regime, up to failure. Figure 6 and Figure 7 illustrate the results of the shear tests conducted on SPM and TPM specimens for constant vertical load $F_v$ = 45.34 N, and lateral displacements $\delta_h \approx \pm 3mm$ (three cycles for each specimen at the displacement rate of 0.8 mm/sec). Such results highlight a light hysteretic response in the examined displacement range (more pronounced in the case of TPM specimens), which is due to micro-plasticity events caused by local stress-concentration, before macro-yielding occurs [Amendola *et al.*, 2016d] [Steele and McEvily, 1976]. The shear tests were repeated at the lateral displacement rate of 2.4 mm/sec, without observing marked differences over the case corresponding to the displacement rate of 0.8 mm/sec [Amendola *et al.*, 2016d].

The experimental tests examined in Ref. [Amendola *et al.*, 2016d] included vertical force vs. vertical displacement compression tests on SPM and TPM specimens. Figure 8 and Figure 9 plot the distributions of the effective compression modulus $E_{eff}$ and the effective shear modulus $G_{eff}$ vs. the the solid volume fraction $\phi$ of SPM and TPM samples, which were extracted from the above shear and compression tests [Amendola *et al.*, 2016d]. The results in Figure 8 and Figure 9 show that quadratic models (solid blue lines) reasonably well fit the experimentally determined scaling laws of $G_{eff}/E_m$ and $E_{eff}/E_m$ with $\phi$ ($E_m$ denoting the Young modulus of Ti6Al4V, cf. Table 2). Such a result confirms the bending-dominated nature of the response exhibited by the SPM and TPM specimens [Meza et al., 2014] [Schaedler et al., 2011].

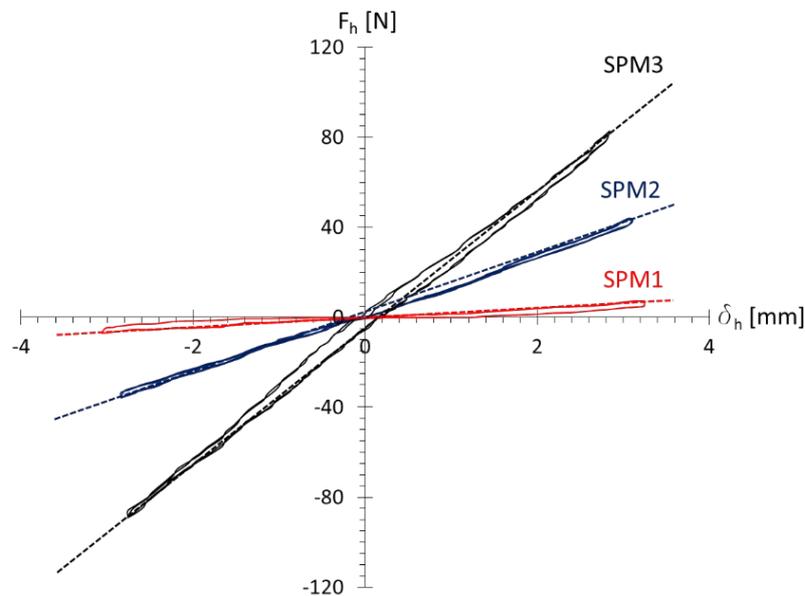

Figure 6 - *Lateral force-displacement curves for SPM specimens.*

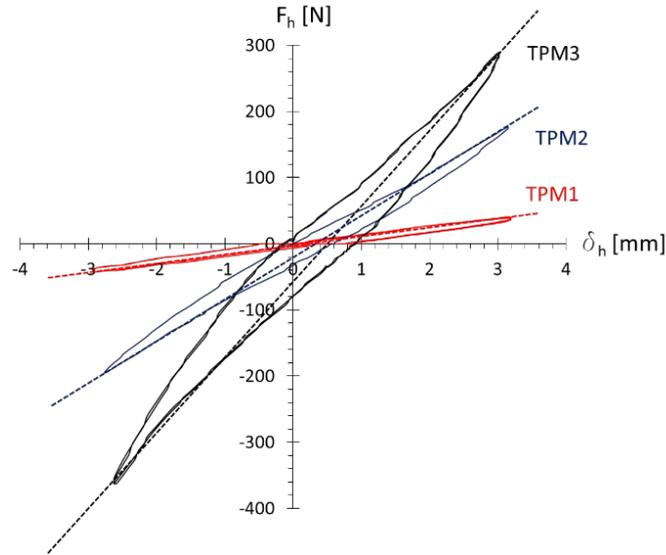

Figure 7 - *Lateral force-displacement curves for TPM specimens.*

The shear and compression tests presented above highlight that $E_{eff}/G_{eff}$ ratio ranges between 7.12 and 7.90 in SPM specimens, and between 2.96 and 4.36 in TPM specimens, with higher values for the specimens with $d=d_3$. Since it theoretically results $E_{eff} \cong G_{eff}$ in pentamode lattices that are not reinforced with stiffening plates [Gurtner and Durand, 2014] [Norris, 2014], we are led to conclude that the considerably high values of such a quantity observed in SPM and TPM specimens are essentially due to the confinement exerted by the stiffening plates against the vertical deformation of the pentamode layers. This response is similar to that of rubber bearings, in which the steel shims restrain the vertical deformation of the rubber layers [Kelly, 1993]. Experimental and numerical results for the elastic moduli of physical models of pentamode lattices have been obtained by Schittny et al. in [Schittny *et al.*, 2013] for 3D printed samples in polymeric materials, which show the same macroscopic aspect ratio of the SPM specimens studied in in Ref. [Amendola et al., 2016d] ($2x2x4$ fcc unit cells). The experimental results presented in [Schittny *et al.*, 2013] estimate the Young modulus $E$ and the shear modulus $G$ of the examined lattices, predicting $E/G$ ratios ranging between 3.67 and 5.52, for varying values of the microscopic aspect ratio $d/a$. The finite element simulations presented in the same work instead predict $E/G$ ratios varying between 4.29 and 4.97. The significant increase of the $E_{eff}/G_{eff}$ ratio observed in the metallic SPM specimens analyzed in in Ref. [Amendola *et al.*, 2016d] over the $E_{eff}/G_{eff}$ ratio of the polymeric samples studied by Schittny et al. is explained by the fact that the SPM specimens show larger and stiffer bases, as compared to the samples analyzed in [Schittny *et al.*, 2013].

We now pass to report the results on the post-elastic response of SPM and TPM samples under cyclic lateral force - displacement tests, which have been presented in Ref. [Amendola *et al.*, 2016d]. Such tests have been conducted up to failure under constant vertical load $F_v$ = 44.45 N, at the lateral displacement rate of 0.8 mm/sec. Figure 10 shows the envelops of the cyclic

force– displacement curves recorded for SPM and TPM specimens, together with the results of monotonic tests conducted on SPM specimens.

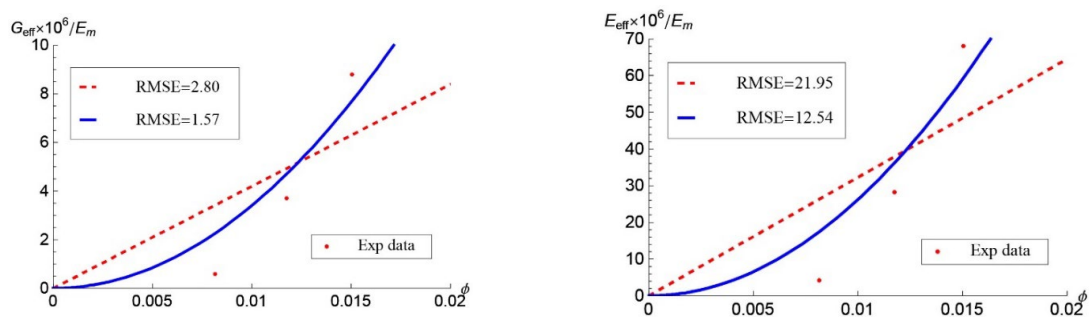

Figure 8 - *Fitting models of the scaling laws of the elastic moduli of SPM specimens with ϕ.*

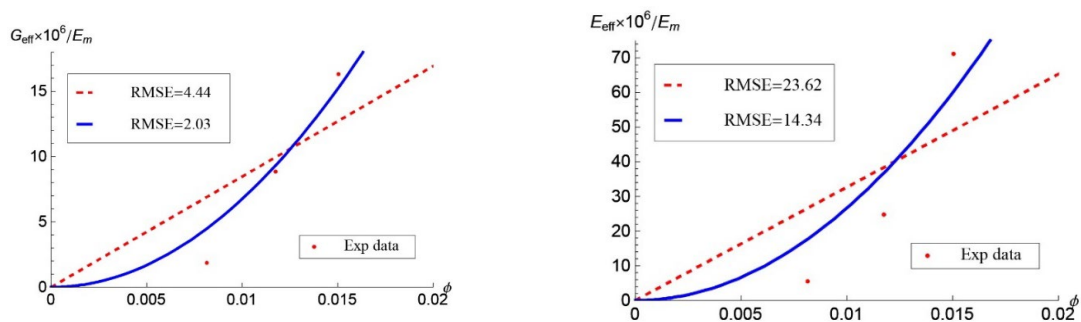

Figure 9 - *Fitting models of the scaling laws of the elastic moduli of TPM specimens with ϕ.*

The initial branch of the curves in Figure 10 is slightly hardening and is followed by a softening branch leading the specimen to failure, due to the progressive collapse of the joints placed along the most stretched rods [Amendola *et al.*, 2016d]. The softening branch is rather steep in the case of SPM specimens, and more flat in the case of TPM specimens.

Figure 11 shows the lateral force-displacement curves recorded for two individual specimens (SPM1 and a TPM3) at controlled ductility, defined as $\mu = (\delta_h - \delta_{h,el})/\delta_{h,el}$ where $\delta_h$ is the current lateral displacement, and $\delta_{h,el}$ is the lateral displacement at the end point of the initial linear branch. The cycles shown in Figure 11 have been obtained by letting the ductility to grow, from the value $\mu = 0.25$ referred to the first cycle, with step 0.25 up to $\mu = 1.00$, and with step 0.50 in the post-elastic cycles up to specimen's failure.

Table 4 shows the Energy Dissipated per Cycle $EDC$; the effective viscous damping $\xi_{eff}$; the post-yield ($K_d$) vs. elastic ($K_e$) stiffness ratio $|K_d - K_e|/K_e$; and characteristic strength $Q_d$ recorded for the SPM1 and TPM3 samples in Ref. [Amendola *et al.*, 2016d]. The data in Table 7 indicate a progressive increase of the dissipated energy with displacement, and values of effective damping that are suitable for seismic isolation devices (25% to 30%).

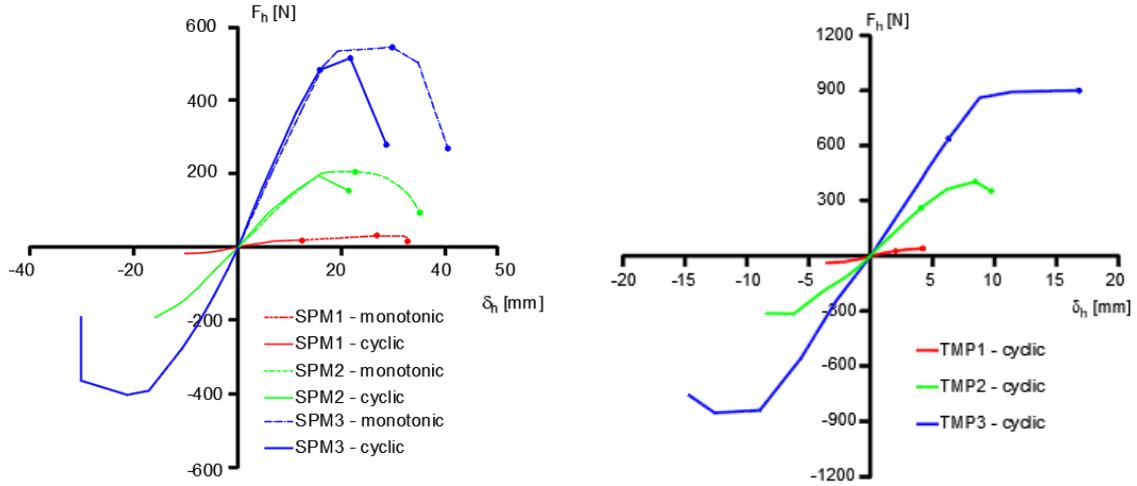

Figure 10 - *Envelops of cyclic and monotonic lateral force-displacement tests up to failure.*

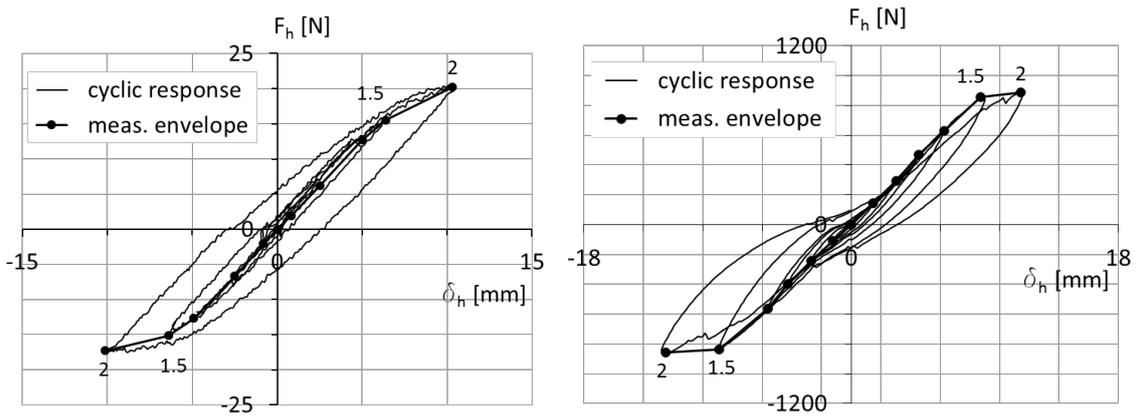

Figure 11 - *Lateral force - displacement responses of specimens SPM1 (left) and TPM3 (right) up to failure (numbers indicate ductility).*

Table 4 - Parameters characterizing the post-elastic response of SPM1 and TPM3 specimens.

|  |  | $EDC$ ($10^3 kN \cdot mm$) | $\xi_{eff}$ (%) | $\|K_d - K_e\|/K_e$ (%) | $Q_d$ (N) |
|---|---|---|---|---|---|
| SPM1 | $\mu_{max} = 1.5$ | 191.44 | 25.11 | 25.73 | 1.52 |
|  | $\mu_{max} = 2.0$ | 406.27 | 31.12 | 57.44 | 5.27 |
| TPM3 | $\mu_{max} = 1.5$ | $12.97 \cdot 10^3$ | 28.02 | 10.07 | 93.12 |
|  | $\mu_{max} = 2.0$ | $18.86 \cdot 10^3$ | 30.12 | 88.73 | 95.01 |

### 4.3 Numerical simulations

A finite element approach to the optimal design by computation of PMBs has been presented in Ref. [Amendola et al., 2016a], with reference to the bending dominated regime of systems equipped with rigid (moment-resisting) connections, considering varying microscopic and macroscopic aspect ratios. The parametric study presented in [Amendola et al., 2016a] considers multilayer *fcc* systems featuring lattice constant $a = 30$ mm, 2x2 unit cells in the horizontal plane, and Ti6Al4V bi-conical struts with diameter at the mid-span $D = 2.72$ mm, as in the 3d-printed samples examined in the previous sections. The optimal values of the microscopic and macroscopic aspect ratios and the number of layers are obtained by maximizing the $E_{\mathit{eff}}/G_{\mathit{eff}}$ ratio. It is worth observing that the macroscopic aspect ratio $H/a$ coincides with the number of layers forming the PMB, since we it is assumed $n_z = 1$ in each layer. The effective values of the compression and shear moduli of PMBs are hereafter denoted $E_{PM}$ and $G_{PM}$, respectively, and are compared to the analogous moduli $E_o$ and $G_o$ of a rubber material ($E_o \cong 4.00$ MPa; $G_o \cong 1.00$ MPa).

The employed finite element model of the generic unit forming the systems analyzed in in Ref. [Amendola et al., 2016a] is illustrated in Figure 12. Such a model has been realized through COMSOL Multiphysics® and is composed of tetrahedral solid elements for both the rods of the pentamode lattices and the stiffening plates (minimum features ranging in between 7% and 20% of the junction size *d*).

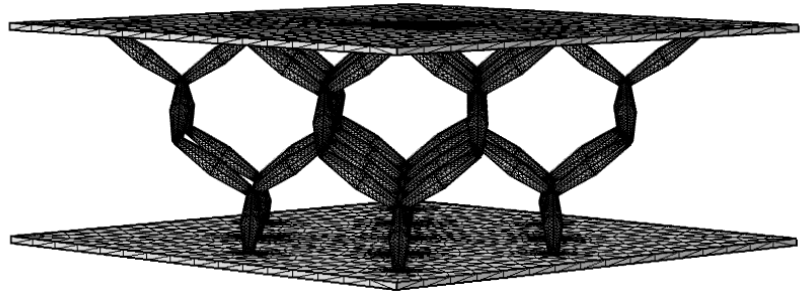

Figure 12 - *Solid finite element model of a confined fcc unit.*

The variation of the effective moduli of finite element models of PMBs with microscopic and macroscopic aspect ratios is illustrated in Figure 13.

One observes from the results in Figure 13 that the largest $E_{PM}/E_0$ and $G_{PM}/G_0$ ratios are reached in single-layer systems equipped with thick junctions. In particular, it results $E_{PM} \cong 70\ E_0$ and $G_{PM} \cong 85\ E_0$ in a single-layer system featuring $d/a=0.09$ (large junction size).

For what concerns the $E_{PM}/G_{PM}$ ratio, we observe from the results in Figure 14 that such a quantity grows both with macroscopic aspect ratio (i.e., the number of layers), and the microscopic aspect ratio (i.e., the size of the junctions). In a five-layer PMB we record $E_{PM} \cong 27\ G_{PM}$ for $d/a=0.09$.

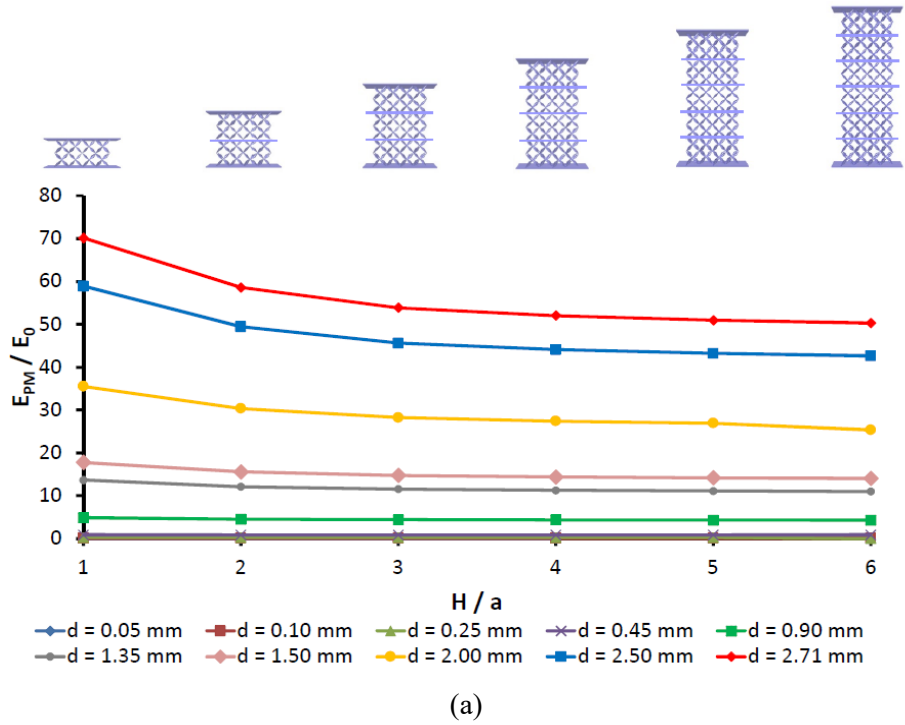

(a)

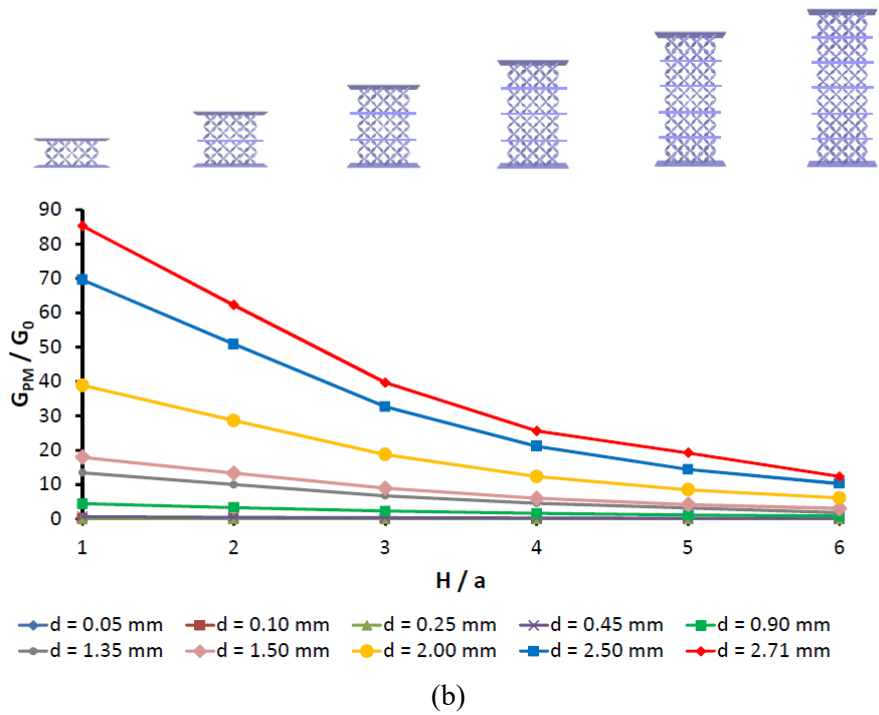

(b)

Figure 13 - *Effects of the tuning of microscopic and macroscopic aspect rations on the effective elastic moduli of multi-layer PMBs.*

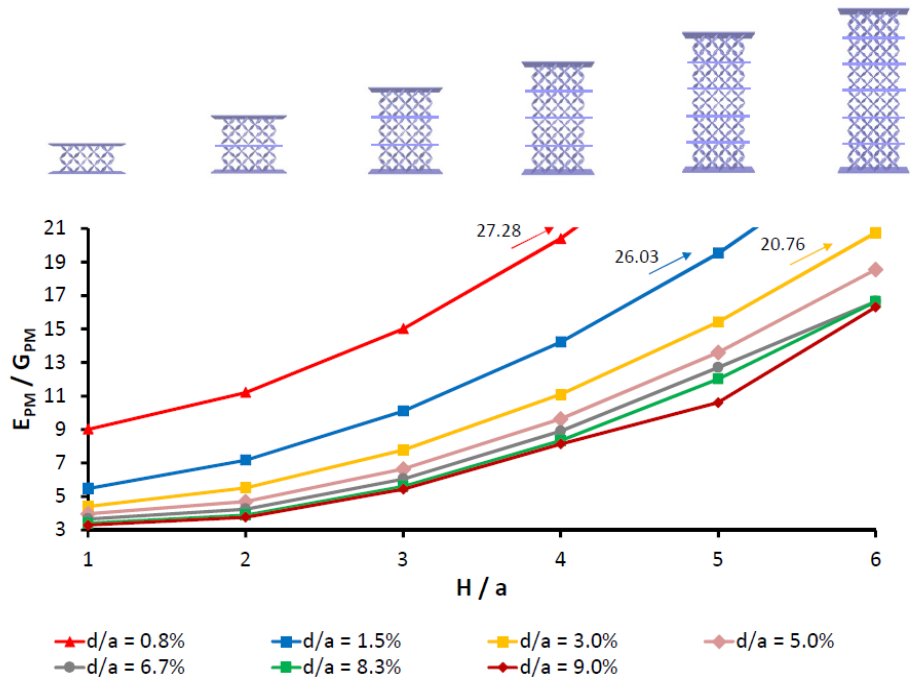

Figure 14 - *Variation of the $E_{PM}/G_{PM}$ ratio with macro-scale and micro-scale aspect ratios of multi-layer PMBs.*

## 5. Concluding Remarks

This paper has reviewed recent literature results on the mechanics of layered structures formed by pentamode layers and stiffening plates, referred to as pentamode bearings (PMBs), and their potential use as innovative base isolation systems.

The results presented in Sects. 3 and 4 regarding the stretching- and bending-dominated regimes of PMBs have pointed out a series of similarities between the mechanical behavior of such structures and that of rubber bearings [Amendola *et al.*, 2016a,b,c,d] [Fraternali and Amendola, 2017]. In both such systems, high values of the effective compression modulus can be achieved due to the confinement of the deformation of the soft (pentamode/rubber) layers that is offered by the stiffening plates. The compression modulus of a sfcc PMB is inversely proportional to the lattice thickness [Fraternali and Amendola, 2017], and one easily recognizes that the role played by the thickness of the rubber pads in rubber bearings is replaced by the area of the cross-sections of the lattice rods in a PMB (cf. Sect. 3).

We have also observed that the ratio between the effective compression modulus and the effective shear modulus of PMBs can get sufficiently high if one suitably designs the microstructure and macrostructure of the pentamode lattices [Amendola *et al.*, 2016a]. It is worth remarking that the mechanical properties of such systems are indeed mainly rooted to the design of the lattice geometry, both at the micro- and the macro-scale, rather than the chemical composition of the material, since such systems respond as mechanical metamaterials [Amendola *et al.*, 2016]. Additionally, an optimized choice of the materials to be employed for the pentamode lattices and the stiffening plates, and a suitable design of the joints (rigid, semi-rigid or hinged) offer additional design opportunities, for what concerns both the effective elastic moduli, and the energy dissipation capacity.

The literature results reviewed by the present paper have highlighted that PMBs can be designed so as to: (i) support the gravity loads at different levels of horizontal displacements; (ii) offer the necessary flexibility in order to increase the period of vibration of the served structure, thus reducing the seismic energy absorption; (iii) reduce damage in structural and nonstructural components (iv); absorb part of the transmitted seismic energy; (v) behave as tension-capable bearings that can bear both compression and tension vertical loads during seismic excitations; (vi) offer sufficiently high stiffness under minor wind loads and low magnitude earthquakes.

The above considerations lead us to conclude positively on the high engineering potential of PMBs for base isolation devices with controlled performance. The values of effective damping observed for the 3d printed samples described in Sect. 4 are aligned with the values of such a quantity observed in common isolation devices [Amendola *et al.*, 2016d]. The same holds regarding the energy dissipated per cycle of energy of the analyzed PMB prototypes, which on the other hand can be markedly increased by reducing material porosity, and/or adding additional dissipative elements within the PMB, such as, e.g., dissipative lead-cores and/or SMA wires [Benzoni and Casarotti, 2009] [Constantinou *et al.*, 2007] [Graesser and Cozzarelli, 1991].

The results presented in Sect. 4 have highlighted that the fabrication of PMBs does not necessarily require heavy industry, and expensive materials, opposed to currently available seismic isolation systems, being possible with ordinary 3-D printers and use of recycled materials. Key advantages like marked tunability, high scalability as well as low manufacturing cost make these devices suitable for applications in developing countries.

In closing, we point out a number of open questions regarding PMBs that suggest directions for future work. The fabrication of such devices through additive manufacturing techniques involving multiple materials (metals, polymers, etc.), and/or ball-joint systems used in space grid technologies awaits attention. Future work is also needed with regards to the experimental testing of real-scale PMBS, the multiscale modeling of plasticity, fracture and damage under large strains [Schmidt *et al.*, 2009][Fraternali *et al.*, 2010][Capecchi *et al.*, 2011][Trovalusci et al., 1998, 2010], and the inclusion of prestressed cables in PMBs, made, e.g., of shape memory alloys, in order to increase the dissipation and re-centering capacities of the device [Naeim and Kelly, 1990] [Graesser and Cozzarelli, 1991].

## Acknowledgements

The authors acknowledge financial support from the Italian Ministry of Education, University and Research (MIUR) under the 'Departments of Excellence' grant L.232/2016.

## References

Amendola, A., Fabbrocino, F., Feo, L., Fraternali, F., Dependence of the mechanical properties of pentamode materials on the lattice microstructure, ECCOMAS Congress 2016 - European Congress on Computational Methods in Applied Sciences and Engineering, 5- 10 JUNE 2016 Crete Island, Greece, No. 6004 (17 pages).

Amendola A., Mascolo I., Orefice A., Benzoni G., Fraternali F. (2018a). Effective stiffness properties of multi-layered pentamode lattices in the stretching-dominated regime, Proceedings of the 16TH International conference of numerical analysis and applied mathematics - ICNAAM 2018.

Amendola, A., Carpentieri, G., Feo, L., Fraternali, F. (2016a). Bending dominated response of layered mechanical metamaterials alternating pentamode lattices and confinement plates. Compos. Struct. **157**, 71-77. ISSN: 0263-8223.

Amendola, A., Benzoni, G., Fraternali, F. (2016b). Non-linear elastic response of layered structures, alternating pentamode lattices and confinement plates. Composite part B, Engineering. ISSN: 1359-8368.

Amendola, A., Carpentieri, G., Feo, L., Fraternali, F. (2016c). Bending dominated response of layered mechanical metamaterials alternating pentamode lattices and confinement plates. Compos. Struct. **151**,71-77.

Amendola, A., Mascolo, I., Benzoni, G. (2018). On the mechanical response of multilayered pentamode lattices equipped with hinged and rigid nodes, PSU research review, In Press.

Amendola, A., Smith, C.J., Goodall, R., Auricchio, F., Feo, L., Benzoni, G., Fraternali, F. (2016d). Experimental response of additively manufactured metallic pentamode materials confined between stiffening plates. Compos. Struct. **142**, 254-262.

Benzoni, G., Casarotti, C. (2009). Effects of Vertical Load, strain rate and cycling on the response of lead-rubber seismic isolators. J. Earthquake Eng. **13**(3), 293-312.

Bückmann, T., Thiel, M., Kadic, M., Schittny, R., Wegener, M. (2014). An elastomechanical unfeelability cloak made of pentamode metamaterials. Nat. Comm. **5**, 4130.

Capecchi, D., Ruta, G., Trovalusci, P. (2011). Voigt and Poincaré's mechanistic-energetic approaches to linear elasticity and suggestions for multiscale modelling. Archive of Applied Mechanics 81(11), 1573-1584.

Chen Y, Liu, X, Hu, G. (2015). Latticed pentamode acoustic cloak. Scientific Reports **5**:15745.

Chilton, J., Space Grid Structures, Oxford, UK, 2000.

Constantinou, M. C., Whittaker, A. S., Kalpakidis, Y., Fenz, D. M., Warn, G. P. (2007). Performance of seismic isolation hardware under service and seismic loading, Technical Report MCEER-07-0012.

European Committee for Standardization. Anti-seismic devices, EN 15129. Brussels, Belgium, 2009.

Fabbrocino, F., Amendola, A. (2017). Discrete-to-continuum approaches to the mechanics of pentamode bearings. Composite Structures **167**, 219-226.

Fabbrocino, F., Amendola, A., Benzoni, G., Fraternali, F. (2016). Seismic application of pentamode lattices. Ingegneria Sismica/International Journal of Earthquake Engineering; 1-2, 62-71.

Fraternali, F., Negri, M, Ortiz, M. (2010). On the Convergence of 3D Free Discontinuity Models in Variational Fracture. Int. J. Fracture **166** (1-2), 3-11.

Fraternali F., Carpentieri G., Montuori R., Amendola A., Benzoni G. (2015). On the use of mechanical metamaterials for innovative seismic isolation systems, COMPDYN 2015 - 5th ECCOMAS Thematic Conference on Computational Methods in Structural Dynamics and Earthquake Engineering, 349-358.

Fraternali, F., Amendola, A. (2017). Mechanical modeling of innovative metamaterials alternating pentamode lattices and confinement plates. Journal of the mechanics and physics of solids **99**, 259-271.

Graesser EJ, Cozzarelli FA. (1991). Shape memory alloys as new materials for aseismic isolation. J Eng Mech ASCE **117**(11):2590-608.

Gurtner, G., Durand, M. (2014). Stiffest elastic networks. Proc. R. Soc. A **470**:20130611. DOI: 10.1098/rspa.2013.0611.

Hernandez-Nava, E., Smith, C. J., Derguti, F., Tammas-Williams, S., Leonard, F., Withers, P. J., Todd, I., Goodall, R. (2015). The effect of density and feature size on mechanical properties of isostructural metallic foams produced by additive manufacturing. Acta Mater. **85**, 387-95.

Huang, Y., Lu, X., Liang, G., Xu, Z. (2016). Pentamodal property and acoustic band gaps of pentamode metamaterials with different cross-section shapes. Phys. Lett. A **380**(13), 1334-1338.

Hutchinson, R.G., Fleck, N.A. (2006). The structural performance of the periodic truss. J. Mech. Phys. Solids **54** (4), 756-782.

Kadic, M., Bückmann, T., Schittny, R., Wegener, M. (2013). On anisotropic versions of three-dimensional pentamode metamaterials. New J. Phys. **15**:023029.

Kelly, JM. (1993). Earthquake-resistant design with rubber. London: Springer-Verlag.

Lomiento, G., Bonessio, N., Benzoni, G. (2013a). Friction model for sliding bearings under seismic excitation. Journal of Earthquake Engineering, 17(8), 1162-1191.

Lomiento, G., Bonessio, N., Benzoni, G. (2013b) Concave sliding isolator's performance under multi-directional excitations. Ingegneria Sismica, International Journal of Earthquake Engineering, 30(3), 17-32.

Martin, A., Kadic, M., Schittny, R., Bückmann, T., Wegener, M. (2010). Phonon band structures of three-dimensional pentamode metamaterials. Phys. Rev. B. **86**, 155116.

Metals Handbook, (2) Properties and selection: nonferrous alloys and special-purpose materials. ASM International, Metals Park, OH, 1990.

Meza, L.R., Das, S., Greer, J.R. (2014). Strong, light weight, and recoverable three-dimensional ceramic nanolattices. Science **6202**(345):1322-1326.

Milton, G.W. (2013). Adaptable nonlinear bimode metamaterials using rigid bars, pivots, and actuators. J. Mech. Phys. Solids. **61**(7), 1561-1568.

Milton, G.W., Cherkaev, A.V. (1995). Which elasticity tensors are realizable? J. Eng. Mater-T. **117**, 4, 483-493.

Naeim, F., Kelly, J.M. (1999). Design of Seismic Isolated Structures: From Theory to Practice, John Wiley & Sons, Inc., Hoboken, NJ, USA.

Norris, A.N. (2014). Mechanics of elastic networks. Proc. R. Soc. A **470**, 20140522.

Pant, D. R., Wijeyewickrema, A. C., ElGawady, M. A. (2013). Appropriate viscous damping for nonlinear time-history analysis of base-isolated reinforced concrete buildings. Earthquake Engng. Struct. Dyn. **42**(4), DOI: 10.1002/eqe.2328.

Schaedler, T.A., Jacobsen, A.J., Torrents, A., Sorensen, A.E., Lian, J., Greer, J.R., Valdevit, L., Carter, W.B. (2011). Ultralight Metallic Microlattices. Science **6058**(334), 962-965.

Schittny, M., Bückmann, T., Kadic, M., Wegener, M. (2013). Elastic measurements on macroscopic three-dimensional pentamode metamaterials. Appl. Phys. Lett. **103**.

Schmidt, B, Fraternali, F., Ortiz, M. (2009) Eigenfracture: An Eigendeformation Approach to Variational Fracture. Multiscale Model. Sim. **7** (3), 1237-1266.

Skinner, R.I., Robinson, W.H., McVerry, G.H. (1993). An introduction to seismic isolation. Wiley.

Steele, R.K., McEvily, A.J. (1976). The high-cycle fatigue behavior of Ti-6Al-4V alloy. Eng Fract Mech. **8**, 31-37.

Trovalusci, P., Augusti, G. (1998). A continuum model with microstructure for materials with flaws and inclusions. Journal De Physique. IV : JP **8**(8), Pr8- 383-Pr8-390.

Trovalusci, P., Varano, V., Rega, G. (2010). A generalized continuum formulation for composite micro cracked materials and wave propagation in a bar. Journal of Applied Mechanics, Transactions ASME **77**(6), 061002.

Van Grunsven W., Hernandez-Nava E., Reilly G.C., Goodall R. (2014). Fabrication and mechanical characterisation of titanium lattices with graded porosity. Metals **4**(3):401-9.